\documentclass[journal, 10pt]{IEEEtranTMBMC}

\normalsize

\newcommand{\vect}[1]{\vec{\boldsymbol{#1}}}
\usepackage{graphicx}
\usepackage{epstopdf}
\usepackage{float}
\usepackage{color}
\usepackage{amsmath}
\usepackage{array}
\usepackage{relsize}
\usepackage{subfigure}
\usepackage{amsmath,amssymb,latexsym}
\usepackage{amsthm}
\theoremstyle{definition}

\usepackage{graphicx}
\usepackage{epstopdf}
\usepackage{float}
\usepackage{color}
\usepackage{amsmath}
\usepackage{array}
\usepackage{relsize}
\usepackage{subfigure}
\usepackage{amsmath,amssymb,latexsym}
\usepackage{amsthm}
\usepackage{multirow}
\theoremstyle{definition}
\usepackage{setspace}
\usepackage{fancyhdr}
\usepackage{adjustbox}
\usepackage{makecell}

\let\baraccent=\= 
\renewcommand{\=}[1]{\stackrel{#1}{=}} 

\theoremstyle{definition}

\theoremstyle{remark}

\ifCLASSINFOpdf
\else
\fi
\begin{document}

 \title{Energy Harvesting and Magneto-Inductive   Communications with Molecular Magnets on Vibrating Graphene  and Biomedical Applications in the Kilohertz to Terahertz Band}

\author{Burhan Gulbahar, \IEEEmembership{Senior Member, IEEE}  

\thanks{$\copyright$  2018  Copyright is owned by IEEE. Accepted for publication in IEEE Transactions on Molecular, Biological and Multi-Scale Communications. Dr. Burhan Gulbahar is with the Department of Electrical and Electronics Engineering and  Applied Research Center of Technology Products, Ozyegin University, Istanbul, 34794, Turkey, (e-mail: burhan.gulbahar@ozyegin.edu.tr).} 
}%

\markboth{Accepted for publication in IEEE Transactions on Molecular, Biological and Multi-Scale Communications}%
{Accepted for publication in IEEE Transactions on Molecular, Biological and   Multi-Scale Communications}

\maketitle

\begin{abstract}
Magneto-inductive (MI)  Terahertz (THz) wireless channels provide significant theoretical performances  for MI communications (MIC) and wireless power transmission (WPT) in nanoscale networks. Energy  harvesting (EH)  and signal generation are critical for  autonomous operation  in challenging medium including biomedical channels. State of the art electromagnetic (EM) vibrational  devices have millimeter dimensions while targeting low frequency EH without any real-time communications. In this article, graphene resonators are combined with single molecule magnets (SMMs) to realize  nanoscale EH, MIC and  WPT with novel modulation methods achieving simultaneous wireless information and  PT  (SWIPT). Unique advantages  of graphene  including  atomic thickness, ultra-low weight, high strain and  resonance frequencies in the Kilohertz to Terahertz band  are combined  with  high and stable  magnetic moments of Terbium(III) bis(phthalocyanine) SMMs.  Numerical  analyses provide tens of nanowatts powers and efficiencies of $10^4 \, W/m^3$ in acoustic and ultrasound frequencies comparable with vibrational EH devices while millimeter wave carrier generation is numerically  analyzed.  Proposed model and communication theoretical analysis  present  a practical framework for challenging applications in the near future by promising simple mechanical design. Applications include nanoscale biomedical tagging including human cells, sensing and communication for diagnosis and treatment,  EH and modulation for autonomous nano-robotics, and  magnetic particle imaging (MPI).
\end{abstract}
 
\begin{IEEEkeywords}
\vspace{-0.0in} 
Single molecule magnet, graphene, energy harvesting, magnetic induction, Terahertz, nanoscale networks, acoustic, nano-robotic, magnetic particle imaging (MPI)
\end{IEEEkeywords}

\IEEEpeerreviewmaketitle

\section{Introduction}
\label{introduction}
 
\IEEEPARstart{M}{agneto-inductive} (MI) communications (MIC) systems have significantly high theoretical performances in  Terahertz (THz) regime   with applications for in-body, e.g., nanoscale biomedical communication networks for in vivo systems, and for on-chip architectures, e.g., high performance three dimensional (3D) on-chip wireless communications, as discussed in \cite{bg1}. In \cite{cr3},   a microscale radio-frequency identification (RFID) system based on near-field magnetic resonance coupling is proposed both theoretically and experimentally for detecting intracellular activities as an experimentally important validation of nanoscale MIC systems for biomedicine.  Noninvasive and autonomous signal generation for MI transceivers  in wideband spectrum including millimeter wave (mmWave) and THz bands is an important challenge. There is currently no practical and low complexity nanoscale solution for generating and modulating high frequency MI carrier signals with simple architectures in combination with energy harvesting (EH) and wireless power transfer (WPT). Single molecule magnets (SMMs)  with quantum phenomena at low temperatures while having large orbital moment and  magnetic stability at  high temperatures are candidates for  simple and high performance building blocks of nanoscale MIC \cite{r15}.   SMMs have important applications including hybrid molecular spintronic devices with graphene in \cite{r15},  high density information storage and quantum computing  \cite{cr4}.   In this article, unique properties of  graphene-SMM hybrid device with SMMs grafted on  vibrating graphene are, for the first time,  utilized  for nanoscale EH, MIC and WPT by exploiting molecular dimensions, atomic thickness layering, ultra-low weight and wideband spectrum in a simple mechanical design while providing nanoscale simultaneous wireless information and power transfer (SWIPT) in the KHz to THz band. 

\begin{figure*}[!t]
\begin{center}
\includegraphics[width=7in]{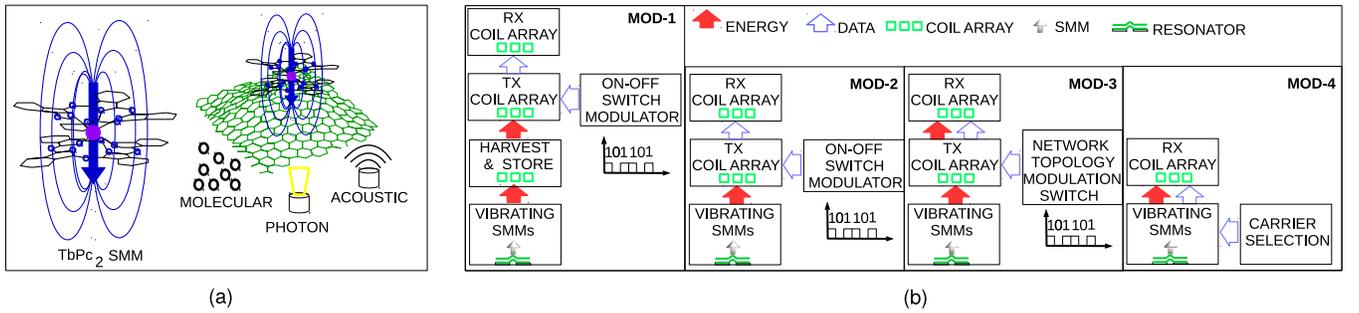} \\
\caption{(a) $\mbox{TbPc}_2$ molecular nanomagnet attached to resonating graphene sheet, and (b)  four different SMM based EH modulation methods with different energy utilization, data modulation and hardware blocks of the coil units.}
\label{figtb1}
\end{center}
\end{figure*}

Graphene nano-electromechanical systems (NEMS) are promising with nanoscale dimensions, high stability  and operation frequencies reaching tens of GHz while electrostatic graphene microphone  covers $20$ Hz to $0.5$ MHz band  \cite{e10}.  Graphene resonators or oscillators are theoretically capable of THz resonance frequencies. They have exceptional properties resulting in high performance such as large Young's modulus (1 TPa for single layer graphene (SLG)), low residual stress and large breaking strength (allowing $25\%$ strain level) for strong and durable devices and ultra-low weight for utilization in challenging medium such as for biological applications \cite{mgr1}. In this article, they are utilized for  MI signal generation  in combination with    grafted SMMs. Graphene-SMM hybrids  preserve magnetic properties of SMM  Terbium(III) bis(phthalocyanine) (Tb$\mbox{Pc}_2$), i.e., a single Tb (III) ion sandwiched between two planar phthalocyanine (Pc) ligands, and graphene,   based on  similar hybrid architectures realized with carbon nanotubes (CNTs) and graphene \cite{r15, e3, e6, cr2}.   Proposed system is applicable for other kinds of SMMs preserving magnetic properties in contact with graphene. Properties of CNT-SMM hybrid are preserved in \cite{e6} with a highly efficient grafting process. In \cite{e3}, static magnetic properties, structural and electronic properties of SMMs on graphene are not modified.  In \cite{cr1}, a super lattice of SMMs ($\mbox{Fe}_4 \mbox{H}$) is formed on graphene with magnetic easy axis of each molecule  oriented perpendicular to the surface and  forming a 2D array of SMMs retaining their bulk magnetic properties.     Nonlinear mechanical properties of CNT are utilized in \cite{e6, cr2} with frequency shift depending on applied magnetic field and magneto-mechanical effects. Graphene-SMM hybrid  is, for the first time, utilized to realize  MIC, EH and WPT simultaneously.

State of the art micro-magnet and micro-coil EH combinations are not designed to produce   wide band and tunable frequency signals while  utilizing lower frequencies with millimeter level areas not suitable for nanoscale communications,  and with efficiencies smaller than $10^5$ $W/m^3$ \cite{e13}.  In \cite{e17}, EH is performed from $330$ MHz and $0.5$ nm vibration amplitude of permanent magnet on  nanoplates with an efficiency on the orders of several $10^3$ $W/m^3$ without detailed analysis of performance scalability with respect to   geometrical and magnetic characteristics of the device.    On the other hand, nanoscale  EH  of different kinds transducing acoustic energy to electricity are future  promising to provide noninvasive and continuous  energy  for autonomous operation \cite{e12}.  However, electrostatic,  piezoelectric, triboelectric, micro-fluid or magnetostrictive types harvest energy from  low frequency ambient vibrations without any target for high frequency signal generation and communications. Besides that, in this article,   a novel method  to produce MI THz signals is proposed with a simple mechanical design as discussed in \cite{bg1}. 

Magnetic resonance imaging (MRI) and  magnetic particle imaging (MPI)  are  limited by the amount and the size of the particles. Resolution with superparamagnetic iron oxide nanoparticles (NPs) (SPIONs) is approximately $1$ mm while promising $300$ $\mu$m resolution with optimized  NPs, hardware and pulses  \cite{e19}. SPIONs are utilized for imaging contrast enhancement, immunoassays, tissue repair, magnetic hyperthermia and drug delivery. However,  finite resolution  is challenging to discriminate cells in the same area. In the proposed article, resonators can be injected into cells or their environments for  carrier based tagging  in  MPI systems.

\begin{table*}[t]
\centering
\caption{Graphene as excitation source or under pressure}
\label{figtabmod}
\footnotesize
\includegraphics[width=7in]{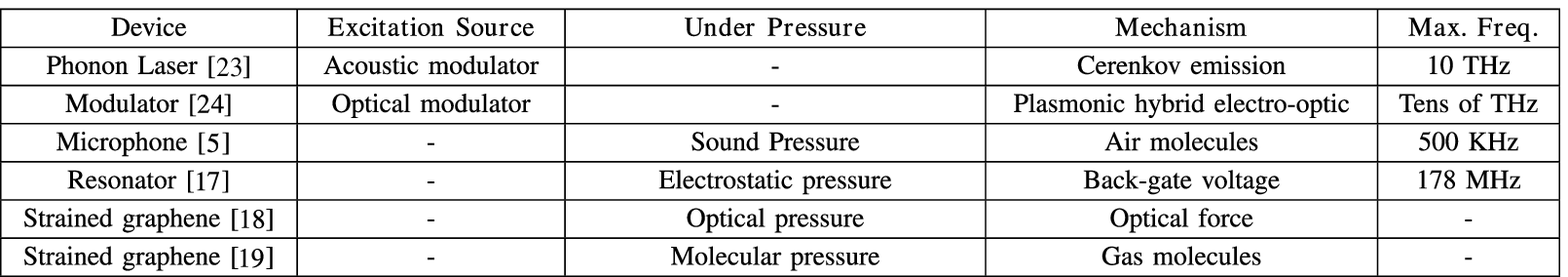}
\end{table*}

The contributions achieved in this article are as follows:
\begin{enumerate} 
\item    SMMs are utilized in a novel nanoscale MIC, EH and WPT application with a special mechanical design  utilizing vibrating graphene and Tb$\mbox{Pc}_2$ type SMMs.  The proposed system has  spatial diversities of multiple SMMs, multiple resonators and multiple layers of resonators, and with computed efficiencies by numerical analyses reaching $10^4 \, W/m^3$ and tens of nanowatts (nWs) peak powers in acoustic and ultrasound frequencies.
\item    Modulation mechanisms are proposed for nanoscale MIC with EH resonators and network topology modulation method introduced in \cite{bg2} for SWIPT. 
\item  Design of a simple, mechanical and EH tagging method  is presented  for   nanoscale and microscale size units.  Future applications include tagging biological structures such as human cells  by utilizing  frequency diversity and hundreds of micrometers communication ranges which can be extended to millimeter and centimeter scales with passive waveguides in \cite{bg1} promising a novel monitoring,   target tracking  and cellular level MPI system.
\item   A system design is proposed  to generate mmWave and THz carrier waves to be utilized in low power and high frequency on-chip and in-body  MIC and WPT  applications while promising  optimizations in future works  such as receiver diversity with multiple carriers. 
\end{enumerate}

The remainder of the paper is organized as follows. In Section \ref{s2}, we present the hybrid graphene-SMM device structure. Then, in Section \ref{s3},  EH and MIC device models are presented for small radius and large radius resonators.   In Section \ref{s4}, the channel models for diversity combining architectures are  presented. Then, in Section \ref{s5},  experimental  challenges are discussed while potential applications are discussed in Section \ref{s5b}.  Numerical analyses  for varying geometrical parameters are performed in \ref{s6}. Finally, Section \ref{conclusion} concludes the paper.

\section{Graphene-SMM Resonator Based Modulation Methods}
\label{s2}
  
Terbium ion molecules  are  attached on graphene  resonator  as shown in Fig. \ref{figtb1}(a)  similar to the architectures proposed for CNTs in \cite{e6} and  \cite{cr2} in terms of the mechanical design. SMMs on vibrating graphene create an oscillating magnetic field which is utilized for inducing currents in a nearby receiver coil based on Faraday's law of induction for MIC, EH and WPT purposes.   In \cite{e6},  a highly efficient grafting process of Tb$\mbox{Pc}_2$ SMMs onto a  CNT NEMS is proposed   while it is  highly sensitive even at a molecular magnet level.   Superlattices  of single atom magnets on graphene are realized with a density of $115$ Tbit/$\mbox{inch}^2$ promising  high density   structures in \cite{e4}. Grafting methods   allow  high density graphene-SMM hybrid design \cite{e6, cr1, e4}.

\begin{table*}[t]
\centering
\caption{SMM based Data and Energy Modulation Methods}
\label{Table_2}
\footnotesize
\includegraphics[width=7in]{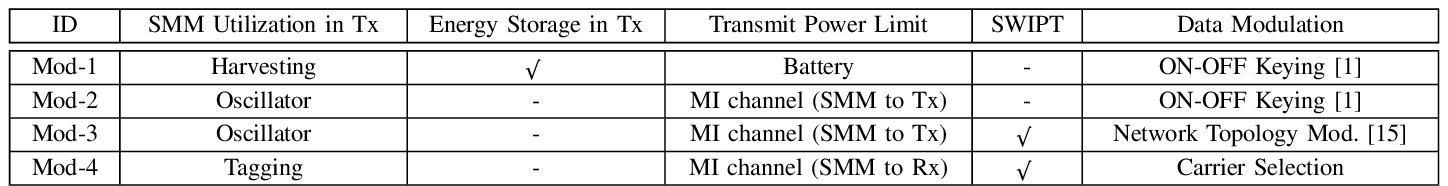}
\end{table*}

\begin{figure*}[!t]
\centering
\includegraphics[width=7in]{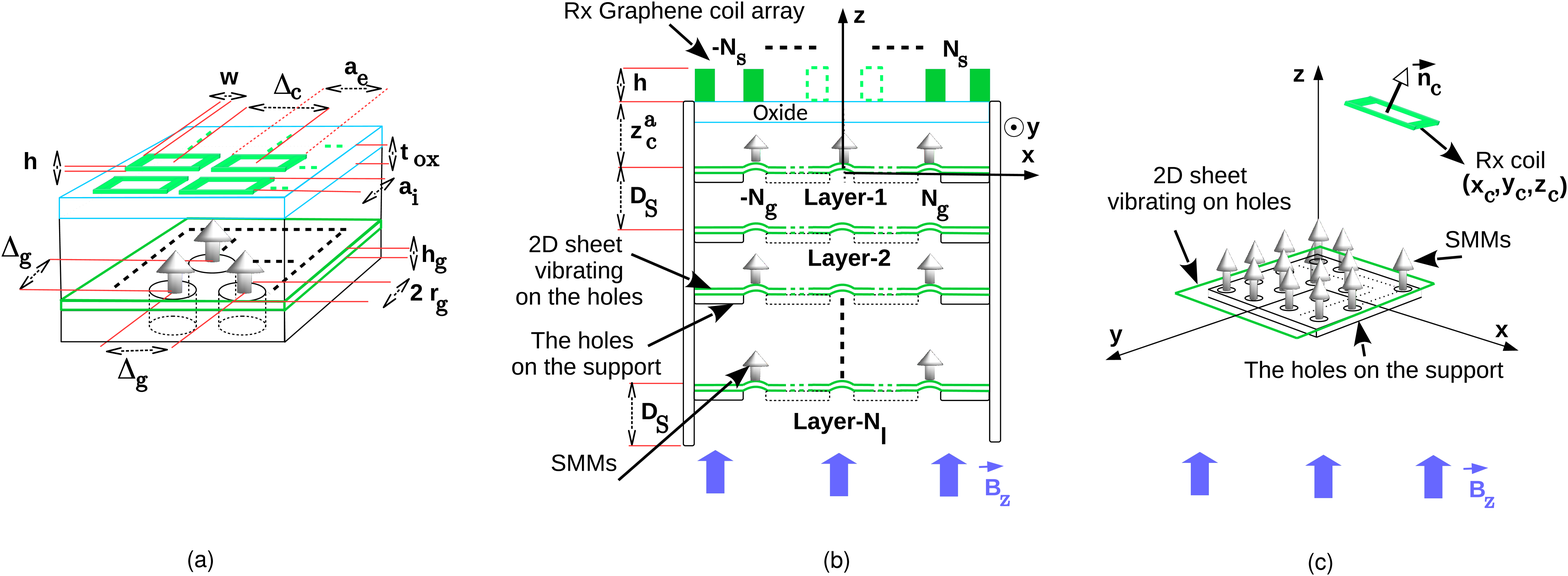}\\
\caption{EH device geometries for the types of (a) SLMO, (b) MLMO, and (c) the system model for SMM based wireless communications channel where the coil resides at $(x_c, y_c, z_c)$ with the normal vector of $\vect{n}_c$. }
\label{Figure_2}
\end{figure*} 

Graphene  resonators are modulated with varying sources, e.g., acoustic \cite{e10}, electrostatic \cite{r11},  optical \cite{opforce},  mechanical strain or molecular forces \cite{grbal1}, as shown in Table \ref{figtabmod}. Examples are provided for both the cases where the graphene is utilized as a  \textit{source}  of acoustic or optical signals   for creating external excitation, and where  it is   modulated  as a  \textit{resonator under pressure}  by  external  excitation. Hundreds of MHz modulation frequencies  are experimentally realized  on graphene resonators  while there is a diversity of methods for specific applications. Tens of THz optical or acoustical modulations are possible by utilizing graphene    sources for  an \textit{all-graphene system design} with simplified manufacturing.     However, special design is necessary to utilize the force produced by graphene sources on graphene resonators as an open issue.    
 
 Theoretical model for resonators is based on experimentally verified  models summarized in \cite{mgr1}.  Resonators are clamped circular plates realized with graphene and  mechanical motion is modeled using plate theory as emphasized in \cite{mgr1, mgr3}.   Harmonic oscillator equation with a forced oscillation of uniform pressure $\, P_g \, \cos(\omega_0 \,   t)$ and force $F_g = P_g \,\pi \, r_g^2 \, \cos(\omega_0 \,   t)$ is utilized where $\omega_0 = 2 \, \pi \, f_0$ is the resonance frequency and $r_g$ is the radius of the resonator. Steady-state resonance amplitude ($d_0$) at the center for the first mode of resonance is approximated by assuming plate model (thick resonators) and  using equations (16) and (22) in \cite{mgr2} as follows:   
\begin{equation}
d_0  = \frac{2 \, \pi \, P \, r_g }{\rho \, h_g \, N_1}    
\, \frac{  \Gamma_{\alpha, r_g} \, \big(1 -  \Lambda_{\alpha, r_g}   \big)}{\omega_0^2 \sqrt{\big(1 - (\frac{\omega}{\omega_0})^2\big)^2 + 4 \, \zeta^2 \, (\frac{\omega}{\omega_0})^2}}           
\label{eqd0}   
\end{equation}
where $\Gamma_{\alpha, r_g} =  \alpha^{-1} \, \big( J_1(\alpha \, r_g)  -  \Lambda_{\alpha, r_g} \, I_1(\alpha \, r_g) \big)$, $\Lambda_{\alpha, r_g} = J_0(\alpha \, r_g) \, / \, I_0(\alpha \, r_g)$, $J_o(.)$ and $I_o(.)$ are the Bessel and modified Bessel functions of first kind and order $o$ for $o \, \in [0, 1]$, respectively, $N_1 = \int_0^{r_g} 2 \, \pi \, r  $ $ W^2(r, \, \theta)  \, dr$, $W(r, \, \theta)$ $ =$ $J_0(\alpha \, r)\,  $ $-$  $\Lambda_{\alpha, r_g} \, I_0(\alpha \, r)$, $\alpha \, r_g$ changes between $2.404$ and $3.196$ transforming from membrane to the plate type as shown in Fig. 2 of \cite{mgr2},  $\sqrt{r_g^2 \, T_g \, / \, D}$ is the transition variable denoted by $k$ in \cite{mgr2},  $D = E \, h_g^3 \, / \, (12 \, (1 \, -  \, v^2))$ is the flexural rigidity, $T_g$ is the pretension in (N/m), $\rho$ is 3D mass density, $v$ is the Poisson's ratio,  $E$ is the Young's Modulus,  $h_g$ is the resonator thickness, $\mu_m$ is the damping coefficient (kg/s) of the medium  and $\zeta$  depending on $\rho$,  $h_g$, $ \omega_0$ and $\mu_m$ \cite{cr5}. Although detailed modeling of $d_0$ is available in \cite{mgr2}, analytical calculations for $d_0$ is simplified with strain based modeling. Assume that deflection at the radial position $r$ is given as $d_0 \, (1 - r^2\, / \,r_g^2)$ with a parabolic approximation of the shape. Then, percentage of strain denoted by $S_m$  is approximated as   $S_m \, / \, 100 =   (2\,d_0^2) \, / \, (3\,r_g^2)   - (2\,d_0^4)\, / \, (5 \,r_g^4)$ for small strain levels. Graphene is highly strong tolerating strain  levels of as much as $25\%$ \cite{mn1}. Performance is measured with respect to a specific level of strain. Detailed analysis for the dependence of strain on the applied force is provided in detail in \cite{cr5} where acoustic forces are utilized in varying medium.   
  
Resonance frequencies ($f_0$)  for SLG and multi-layer graphene (MLG) are calculated with experimentally verified plate and membrane models in (8) and (9)  in  \cite{mgr1}  as follows: 
\begin{eqnarray}
f_{0, SLG} =  \frac{2.404}{2 \pi r_g}\sqrt{\frac{T_g}{h_g \, \rho}}\hspace{3.9cm}&& \\
f_{0, MLG} = \sqrt{   f_{0, SLG}^2 + \Bigg( \frac{10.21}{4\, \pi}  \sqrt{ \frac{E}{3\,\rho (1\,-\,v^2)}  } \frac{h_g}{r_g^2}\Bigg)^2} &&                
\end{eqnarray}
It  is assumed that magnetic moments of SMMs are aligned in parallel with the external magnetic field leading to zero magnetic torque compared with magnetometer structures in \cite{e6, cr2}. Therefore, SMMs behave as loads on resonating layer without changing magnetic and structural properties of SMM and graphene with methods in \cite{ e3, e6, cr2, cr1,   e4}. The change in $f_0$ of graphene layer is approximated  in \cite{mgr3} by   $\vert \partial f \vert= \,f_0 \,\vert \partial m \vert/ \, (2 \,m )$ where $m$ is the resonator mass and $\vert \partial m \vert$ is the load leading to $\vert \partial f \vert= f_0 \,  m_{M, SMM}  \, / \, (2 \,m_G)$ where $m_G = \rho \, h_g \, \pi \, r_g^2$ and $m_{M, SMM} = N_{SMM} \, m_{SMM}$ are the weights of  resonator and  SMMs grafted on it, respectively, $N_{SMM}$ is the number of SMMs and $m_{SMM}$ is the weight of each SMM. $N_{SMM}$ is chosen small for the minor effects on resonance, i.e., $m_{M, SMM}  \, = \,\eta_{SMM} \, m_G$ where $\eta_{SMM} \ll  1$. 
     
In Table \ref{Table_2}  and Fig. \ref{figtb1}(b), four different  modulation  methods are proposed where energy is harvested from graphene-SMM resonators. Mod-1 stores harvested energy and utilizes  later. Microscale coils in \cite{bg1} utilize 1 $\mu$W transmit (Tx) power to achieve THz communications and this energy can be stored in several minutes with the efficiency of $\approx 1$ nW, resonator radius of $50 \, \mu$m and coil radius of 5 $\mu$m as discussed in Section \ref{s6}. Mod-2 utilizes real-time simultaneous energy harvesting and data modulation by switching the coil ON and OFF   and  modulating harvested carrier signal. Mod-3 utilizes network topology modulation defined in \cite{bg2} for SWIPT by modulating array of active coils and by utilizing real-time harvested energy without any signal modulation. Mod-4 transmits carrier signal directly without any Tx coil while receiver (Rx) estimates the frequency    for tagging purposes. In this study,  Mod-1 for EH  and Mod-4 for   tagging are analyzed while the others  are open issues.   Next, SMM magnetic moment modeling is presented by emphasizing  the relation with the graphene surface and  relaxation.
   
\subsection{ SMM Spin Orientation and Relaxation Modeling}
\label{s2s2}

In this article, it is assumed that the proposed device operates at room temperature. Magnetic moments of SMMs are assumed to be aligned with an external magnetic field while optimum system design and engineering of external fields and SMM orientations are open issues. In addition, it is assumed that magnetic easy axis of each SMM molecule is oriented perpendicular to the graphene vibrator as discussed in \cite{cr1} and there is no magnetic torque applied on graphene due to misalignment between SMM moment and external field as in \cite{e6, cr2}. The effects of misalignments resulting in frequency shift  are left as future works. 

On the other hand, graphene-SMM hybrid vibration  promises relaxation based analysis similar to  MPI \cite{e19}.   Relaxation time at high temperatures is given by Arrhenius equation with $\tau_{N} = \tau_{0} e^{\Delta   /   k_B \, T} $ where $\Delta$ is the anisotropy barrier given as 566 $\mbox{cm}^{-1}$ for $\mbox{TbPc}_2$  SMM and assuming $\tau_0 = 10^{-9}$ seconds \cite{new_t1}, and $k_B$ is approximately  $0.695 \, \mbox{cm}^{-1}$/K.  Room temperature ($k_B \, T \,  \approx \, 200\, \mbox{cm}^{-1}$) results in $\tau_N \approx 15.1  \, $ ns with $1 \, / \, \tau_N \approx 66$ MHz. Future studies utilizing the proposed resonator at room temperature are limited by relaxation time on the order of tens of nanoseconds. Mechanisms to exploit the proposed system for MPI purposes are left as future works.  Next, modulator device architecture is modeled by providing analytical formulation of inductive channel. 

\section{  Magneto-inductive Modulator Modeling}
\label{s3}

Oscillators are placed on a grid of holes, and coils induced by vibrations are placed in a multi-receiver architecture. Device geometries for EH purposes with single layer (SL) multiple oscillator (MO) and multiple layer (ML) MO  abbreviated as SLMO and MLMO, respectively, are shown in Figs. \ref{Figure_2}(a) and (b), respectively. Single resonator case is denoted with SLSO. EH device is realized with closely and vertically  separated, and parallel oriented  hybrid structures while   MIC design is shown in Fig. \ref{Figure_2}(c) where Rx coil resides at $(x_c, y_c, z_c)$ with the normal vector  $\vect{n}_c$.  The center of resonators is  positioned at the origin of the reference system.  $\Delta_g  = (k_{\Delta g}  \, + \, 2)\, r_g$  and $\Delta_c = (k_{\Delta c} \, + \, 2)\, a_e \, / \, 2$ denote the central distances of SOs and  coils, respectively, where  $k_{\Delta g}$  and $k_{\Delta c}$ are constant separation factors. Inner and outside lengths of  coils are $a_i$ and $a_e$, respectively, where they are determined with respect to the width ($w$) and the height ($h$) of the coil wire with $a_i = 2 \times k_w \times w$, $a_e =a_i \, + \, 2 \times w$, $k_w$ is a constant and $w = h$ with a square cross section.  Coil resides on an oxide substrate with a thickness of $t_{ox}$   with  circuit model presented in Appendix as a simple device model to characterize the approximate performance.  

\begin{figure}[!t]
\centering
\includegraphics[width=3.5in]{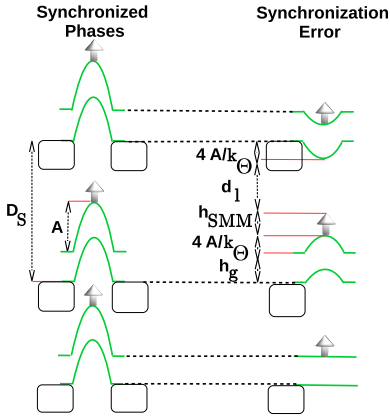}\\
\caption{The effect of the synchronization error in MLMO resonators and the calculation of $D_S$. }
\label{Figure_3}
\end{figure}

Distance of the first layer of SMM oscillators to the bottom of the coil (neglecting oxide thickness much smaller than $h$) is denoted with $z_c^a$ where  z-axis position of the central coordinate of coil is found by $z_c = z_c^a \, + \, h/2$. $D_S$ denotes the thickness of a SL calculated by allowing phase synchronization errors between different layers as shown in Fig. \ref{Figure_3}. The maximum number of layers is assumed to be limited by $N_l$. The layer thickness is assumed to be equivalent to $D_{S} \equiv   d_{l} +  h_{M} + h_g + 8 \, A \, / \, k_{\theta}$ where $d_l$ is the inter-layer separation distance, $A$ is the vibration amplitude, $h_{M}$ denotes the height of the SMM cluster on the resonator (on the order of nms) and $k_{\theta} \gg 1$ is the tolerance factor to the phase errors due to excitation sources or other mechanical effects. It is assumed that the total thickness of the layers is limited by $D_{T}^{max} \equiv \lambda_e/k_{\theta}$ where $\lambda_e = v_e \, / \, f_0$ is the excitation signal wavelength and   $v_e$  is the velocity of the excitation signal. The worst  synchronization error is tolerated with a distance between the layers, i.e., $1 \, / \, k_{\theta}$ multiplied by the total distance that the SMMs in two consecutive layers vibrate which is $2 \times 4 \, \times A$. Then,   $N_{l}   \leq     D_{T}^{max} \, / \, D_{S}$
where $2 \times N_s + 1$ and $2 \times N_g + 1$ denote the numbers of Rx coils and resonators   in a row, respectively, with a square grid placement as shown in Figs. \ref{Figure_2}(a) and (b). Static magnetic field is assumed to be in the $z$ axis with the value $\vect{B}_z$ while a scanning device can be utilized to find the best orientation in practice.   Next, architectures denoted as Sch-1 and Sch-2 are described  improving the geometry in Fig. \ref{Figure_2}(b).

\subsection{Scheme-1 Device Geometry}
\label{s3s1}

\begin{figure*}[!t]
\centering
\includegraphics[width=7in]{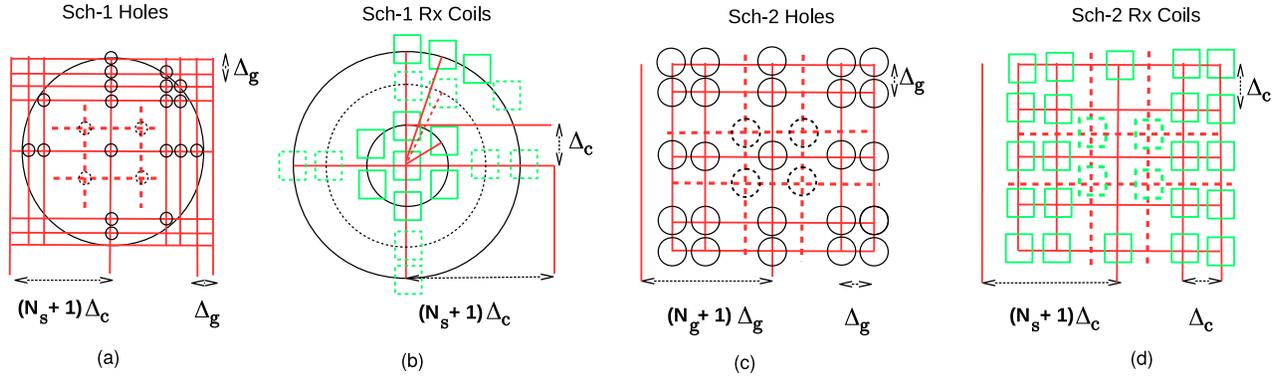}\\
\caption{The geometries of (a) the holes and (b) the coils in Sch-1, and (c) the holes and (d) and the coils in Sch-2 device types.}
\label{Figure_4}
\end{figure*} 
 
In Sch-1, the  dimension of each SO is assumed to be very small compared with the dimension of the coil such that a large  number of resonators is placed under the coil especially for EH or carrier  generation.  Geometries of  holes and coils are shown in Figs. \ref{Figure_4}(a) and (b), respectively, where the total area is a circular region with radius determined by $(N_s \, + 1)\, \Delta_c$,   and $N_s$    this time denotes the number of circularly placed layers of coils  as different from Fig. \ref{Figure_2}(b). Coils are more compactly placed compared with the square grid structure.  Assuming rotational symmetry of the induction,   induced voltage is calculated by grouping SOs with respect to the distance to the coil centers. This lowers the  complexity  due to double integrations in formulations in  Section  \ref{s4} for each pair of the coils and the resonators. Total area of  resonators is compatible with the total area of  coils such that total number of SOs  is approximated by $\pi  (N_s \, + \, 1)^2 \, \Delta_c^2 \, / \, \Delta_g^2$. The angle between  radial lines connecting two neighbor coils on $j$th circle with the radius of $j \, \Delta_c$ is set to $\Psi(j) = \cos^{-1} (1 - 0.5/(j-1)^2 )$ for $j \geq 2$, and $j = 1$ denotes a single coil case with $\Psi(1) \equiv 2 \, \pi$ and $N_s= 0$.  Total number of coils is given approximately by $N_{coil}  = \sum_{i=1}^{N_s+1}  2 \, \pi / \Psi(i)$.     

\subsection{Scheme-2 Device Geometry}
\label{s3s2}

In Sch-2, larger resonators comparable with coils are  utilized, and a square grid based placement is realized as shown in Figs. \ref{Figure_4}(c) and (d). Resonators and coils are placed at  positions $[m \, \Delta_g, n \Delta_g, 0]$ for $m, n \, \in [-N_g, N_g]$ and  $[ m \, \Delta_c, n \Delta_c, z_c]$ for $m, n \, \in [-N_s, N_s]$, respectively, with  a total of $(2 \, N_g + 1)^2$ resonators and $(2 \, N_s+ 1)^2$ coils, respectively. $N_g$ is chosen in compatible with  total area of  coil array with the formulation for rounded integer values of $N_g \approx  (N_s  \, + \, 1/2) \, \Delta_c \, / \, \Delta_g - 1/2$.  Next,  induction channels between SMMs and  coils are modeled.

\section{Wireless Channel Model}
\label{s4}

MI channels for SLSO, SLMO and MLMO cases are modeled consecutively. Magnetic dipole moment  ($\vect{m}$) amplitude  of Tb ion is calculated by  $ g_J \, \mu_B \, J$ where $g_J = 1.326$ is Lande g-factor, $J =6$ is the total angular momentum quantum number and $\mu_B = 9.27 \times 10^{-24}$ J/T is Bohr magneton \cite{e6}. Next, Faraday's law of induction is used for produced  dipole field.

\subsection{Single SMM to Single Coil Wireless Channel}
\label{s4s1}
    
Magnetic induction of a single SMM dipole on a rectangular coil is modeled as follows. Flux $\Phi_{\widetilde{z}, u}$ through a square area of side lengths $2 \, u$  at  $\widetilde{z}$ is  approximated as  $\Phi_{\widetilde{z}, u} = \int_{-u}^{u}\int_{-u}^{u} \vect{B} \cdot\vect{n}_c \,  d\widetilde{x} \, d\widetilde{y}$ where $(\widetilde{x}, \widetilde{y}, \widetilde{z})$  is the local coordinate system of the square with the coinciding center,  sides of the rectangle  are parallel with $\widetilde{x}$  and $\widetilde{y}$ axes, $\vect{B}$ is the magnetic field along the area of  region, and $\vect{n}_c$ is the normal vector of the square area as shown in Fig. \ref{Figure_2}(c). Assume that  coil center is at $\vect{s}_c = [x_c, \, y_c, \, z_c]$.  Then,   generated field due to a single SMM of the moment $\vect{m}$ is found by modifying the equations (2) and (3)  in \cite{noncoupling}  with the randomly oriented coil as  follows:
\begin{equation}
\label{eqBfield}
\vect{B}  = \frac{ \mu _0 \, \vert \vect{m} \vert  }{4 \pi } \,  \frac{ 3 \,  x \, z \,  \vect{e}_x \, + 3 \, y \, z \, \vect{e}_y  -\left(r^2 \,- \,2 \, z^2\right) \vect{e}_z}{\left(r^2 \, + \, z^2\right)^{5/2}}
\end{equation}
where $\vect{e}_{x}$, $\vect{e}_{y}$ and $\vect{e}_{z}$ are  unit vectors of the Cartesian coordinate system $(x, \, y, \,z)$ that SMM resides as shown in Fig. \ref{Figure_2}(c), $\lbrace r,\,x, \, y, \,z \rbrace$ are functions of $(\widetilde{x}, \widetilde{y}, \widetilde{z})$ such that  $r  \, = \sqrt{  x^2 \, + \, y^2}$ and $\left[ x\, y \, z \right]^T = \vect{s}_c \, + \, \widetilde{x}\, \vect{e}_{\widetilde{x}} \, + \,   \widetilde{y} \, \vect{e}_{\widetilde{y}}  + \, \tilde{z} \,  \vect{n}_c$ where $\vect{e}_{\widetilde{x}}$  and $\vect{e}_{\widetilde{y}}$ are unit vectors on the plane of the square coil  in local coordinate system. If  SMM vibration has a sinusoidal form then,  $z_c(t)  =  z_c  \, + \, A \,  \sin (\omega_{0} \, t)$.  Then,  total flux through coil becomes $\Phi(t) = F_c \, N_T\, (1\, / \, A_c) \, \int_{-h\,/\,2}^{ h\,/\,2 } \int_{a_i \, / 2}^{a_e \, / \, 2} \Phi_{\widetilde{z}, u} \, du \, d\widetilde{z}$ where $F_c = N_T \, A_s \, / A_c \, $ is the fill-factor, $A_s$ and $A_c \, = \, w \times h$  are the areas of the coil cross-sections of single turn and total coil device, respectively, and $N_T$ is the number of turns. Induced voltage  $V_{S}(t)  \equiv -d \Phi(t)  \,/ \, dt$ on the coil is found by using $-d\vect{B}\,/ \, dt $ as follows: 
\begin{eqnarray}
\label{app2eq2}
V_{S}(t)   =   \overline{m} \, \dot{z} \, \int _{a_i\, / \, 2}^{a_e\, / \, 2} \int _{-\frac{h}{2}}^{\frac{h}{2}}  \dfrac{\partial  \Phi_{\widetilde{z}, u}^n }{\partial z} d\widetilde{z}\,  du 
\end{eqnarray}
where  $ \Phi_{\widetilde{z}, u}^n$ is the normalized flux with unity $\, \vert \vect{m}  \vert$ in  (\ref{eqBfield}), $\overline{m} \equiv  \mu _0 \, \vert \vect{m}  \vert \,  F_c \, N_T \, / \, (4 \, \pi \,  A_c )$, $\dot{z}  = \dot{z_c}(t)  = -d z  \,/ \, dt$   and  $\mu_0 = 4 \,\pi\, 10^{-7}$ ( $\mbox{T}\times \mbox{m}/\mbox{A}$) is the vacuum permeability. Expression is simplified with assumptions about the coil orientation and that  rotating rectangular coil around its central axis does not change the induction significantly.  If $\vect{n}_c =  \vect{e}_z$, $y_c = 0$, $r_c = x_c$,  $\vect{e}_{\widetilde{x}} = [1\, 0 \, 0]$ and $\vect{e}_{\widetilde{y}} = [ 0 \, 1 \, 0]$, then $\Phi_{\widetilde{z}, u}^n$ becomes $ 2 \, u \, (\zeta_{+} \, \chi_{+} \, + \, \zeta_{-} \, \chi_{-}) \, / \, (\zeta_{-} \, \zeta_{+} \, \varrho )$ where $\zeta_{\pm} = \sqrt{ \Upsilon_{\pm}  \,+ \, u^2}$,  $\Upsilon_{\pm} = \gamma_{\pm} \,+ \beta $, $\beta =  ( z_c + \widetilde{z})^2$, $\gamma_{\pm} = (u \pm x_c)^2$, $\chi_{\pm} = \Upsilon_{\pm} (u  \, \mp \, x_c) \left(\gamma_{\mp}+2 \beta +u^2\right)$ and $\varrho = \Upsilon_{-} \Upsilon_{+} \left(\beta + u^2\right)$.  Then, $  (- 1 \,/ \, 2 u) \, \partial  \Phi_{\widetilde{z}, u}^n  \, / \,  \partial z $ is found as follows: 
 \begin{equation}
  \begin{aligned}
 \frac{\varrho\,\zeta_{+}^2 \chi_{+}  \frac{\partial \zeta_{-}}{\partial z_c}+\zeta_{-} \zeta_{+} \frac{\partial \varrho}{\partial z_c} (\zeta_{-} \chi_{-}+\zeta_{+} \chi_{+})}{\zeta_{-}^2  \, \zeta_{+}^2  \, \varrho^2}  &  \\
+ \,\frac{ \varrho  \, \zeta_{-} \bigg(\zeta_{-} \chi_{-} \frac{\partial \zeta_{+}}{\partial z_c}-\zeta_{+} \left(\zeta_{-} \frac{\partial \chi_{-}}{\partial z_c}+\zeta_{+} \frac{\partial \chi_{+}}{\partial z_c}\right)\bigg)}{\zeta_{-}^2 \, \zeta_{+}^2 \, \varrho^2}   & 
\end{aligned}
  \end{equation}
where  derivatives are  
$\partial \zeta_{\pm} \,  / \, \partial z_c    $ $=$  $   \sqrt{\beta} \, / \, ( \sqrt{\Upsilon_{\pm} + u^2})$,  $ \partial \chi_{\pm} \,  / \, \partial z_c  $ $ =$  $  4 \sqrt{\gamma_{\mp}  \beta} \,\Upsilon_{\pm}$ $+$ $2 \sqrt{\gamma_{\mp} \, \beta} $ $\times \left(\gamma_{\mp}+2 \beta + u^2\right)$ and
$ \partial \varrho \, / \, \partial z_c   $ $=$ $   2 \sqrt{\beta} \big( \left(\beta + u^2\right) (\Upsilon_{+} + \Upsilon_{-} )+  \Upsilon_{-}\Upsilon_{+}  \big)$. The maximum amplitude of the generated voltage is calculated by inserting $\dot{z} = A \, \cos (\omega_{0} \, t) \, \omega_{0} $ into  (\ref{app2eq2}) resulting in the value denoted by $\vert V_{i} \vert \equiv \mbox{max} \lbrace \vert V_{S}(t) \vert \rbrace$.       If  SMMs    are positioned at the center of the coordinate frame on the xy plane, then $V_{i}$ on the coil at $(x_c, y_c, z_c)$   is given  as follows:
\begin{eqnarray}
\label{eq2}
  V_{i}    =   A\, \omega_0 \, \overline{m} \, \int _{a_i \, / \, 2}^{a_e \,  / \, 2} \int _{-\frac{h}{2}}^{\frac{h}{2}}  \dfrac{\partial  \Phi_{\widetilde{z}, u}^n }{\partial z} d\widetilde{z}\,  du   
\end{eqnarray}
while depending on $z_c$ and $r_c = \sqrt{x_c^2 \, + \, y_c^2}$. Multiple SMMs are used for diversity next. 
 
\subsection{Diversity Combining with  Multiple SMMs }
\label{s4s2}

Induced voltage is improved   with collective moments of a group of SMMs.  Their relative positions are assumed the same since they are placed in an area of radius $k_{r} \, r_g $ where $k_r \ll 1$   for large $r_g$ while SMMs placed in a compact manner to the center for $r_g$ near several nanometers. As discussed in   Section \ref{s2}, ratio of the total weight of the SMMs to the weight of 2D layer is $\eta_{SMM}$  with an overall effect of  the frequency shift to $(1 - \eta_{SMM} \, / \, 2) \times f_0$.   Then, the height of the total group of SMMs is roughly estimated as  $h_M =  h_{SMM} \, \mbox{max} \big \lbrace    N_{SMM}\, / \, \big(   \pi \,(k_{r} \, r_g)^2 \, / \, a_{SMM}^2  \big), \,1  \big \rbrace$ where  $a_{SMM}$  is the side length of the placement square for a single SMM and $h_{SMM}$ is  its height. The resulting $V_{i}$ is denoted by $V_{z_c, \, r_c}^{SLSO}$. Next,  multiple SOs are utilized to form SLMO. 
   
\subsection{Diversity Combining with Multiple Oscillators}
\label{s4s3}

\begin{figure}[!t]
\centering
\includegraphics[width=3.5in]{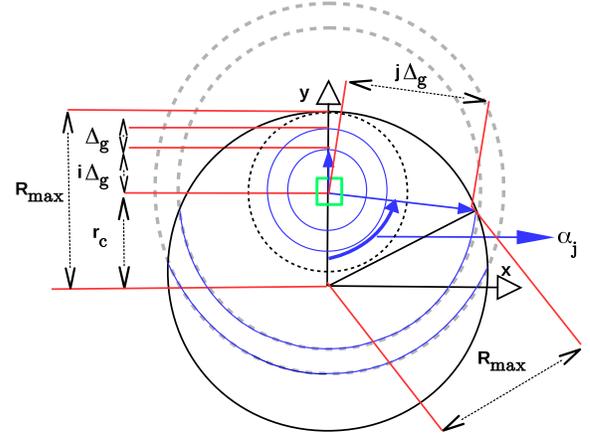}\\
\caption{The geometry for the calculation of the induced voltage in Sch-1 case for $r_c \leq R_{max}$ case. }
\label{Figure_5}
\end{figure}  

Total  $V_{i}$ for  Sch-1 on a coil at radial and vertical distances $r_c$  and $z_c$, respectively,  is denoted by $V_{z_c, r_c}^{SLMOc}$ where  calculation method and geometry are shown in Fig. \ref{Figure_5} for $r_c \leq R_{max}$ with the  arcs for two different regions while  radius of  total  area is $R_{max} \equiv (N_s + 1) \, \Delta_c$,  $r_d \equiv  R_{max} - r_c$ and each SO with an area of $\approx \Delta_g^2$.  It is approximated for $r_c \leq R_{max}$ as follows:
 \begin{equation}
  \begin{aligned}
  V_{z_c, r_c}^{SLMOc} \approx \sum_{i=0}^{  \lfloor  r_d / \Delta_g \rfloor }  \Gamma_n \, C(i) \, V_{z_c, i \, \Delta_g}^{SLSO}  \,    \hspace{2.8cm}  &   \\
+   \,  \sum_{ j =  \lfloor  r_d / \Delta_g \rfloor + 1}^{ \lfloor  (2 \, R_{max} \, - \,  r_d) / \Delta_g \rfloor}  (\Gamma_n  \,  \alpha_j / \pi) \,    C( j) \, V_{z_c,  j \, \Delta_g}^{SLSO}  &  
  \end{aligned}
  \label{vieq5}
\end{equation}
where $ \lfloor . \rfloor$ is the floor function and $\alpha_j =  \cos^{-1}\big(- (R_{max}^2 - r_c^2  - j^2 \Delta_g^2 )/(2 \, r_c \, j \, \Delta_g) \big)$ gives the proportional angle of the arc to calculate its area while if $r_c > R_{max}$, the approximation becomes     $V_{z_c, r_c}^{SLMOc}  \approx \sum_{i = i_s}^{i_e }     \, (\Gamma_n\, \alpha_i / \pi) \,   C(i) \, V_{z_c, i \, \Delta_g}^{SLSO}$ where   $i_s = \lfloor (r_c - R_{max}) \, / \,(\Delta_g)   \rfloor +1$, $i_e =  \lfloor (r_c + R_{max}) \, / \, (\Delta_g) \rfloor$,  and $C(i) = \pi \big(  (i+1)^2 - i^2 \big)$ is the approximate number of SOs in the arc area where radial distance of the coil to the resonators in the circular area between the indices $i$ and $i+1$ becomes $\approx i \Delta_g$. Result is normalized by $ \Gamma_n \equiv (\pi R_{max}^2 \, / \, \Delta_g^2)  \, / \, \sum_i C(i)$.   In Sch-2,  $V_i$ denoted by $V_{z_c, i_x, i_y}^{SLMOr}$  in the coil at $[i_x \, \Delta_c, i_y \, \Delta_c, z_c]$  is approximated as follows:
\begin{eqnarray}
\label{vieq1}
V_{z_c, i_x, i_y}^{SLMOr} \approx \sum_{m=-N_g}^{N_g} \sum_{n=-N_g}^{N_g} V_{z_c, r_c(i_x, i_y, m, n)}^{SLSO}   &&    
\end{eqnarray}
where  $r_c(i_x, i_y, m, n) = \sqrt{ \varsigma_{m, i_x}^2 \, + \, \varsigma_{n, i_y}^2  }$  between  each SO and the coil, and $\varsigma_{i, j} \equiv i \, \Delta_g - j \, \Delta_c$. 
  
\subsection{Diversity Combining with Multiple Layers}
\label{s4s4}

Total voltage values for Sch-1  and  Sch-2 cases are given by the following, respectively:
\begin{eqnarray}
\label{vieq6}
V_{ z_c, r_c}^{MLMOc}  = \sum_{i = 1}^{N_{l}^{*}}  V_{z_c + (i-1) \, D_S, r_c}^{SLMOc} && \\
V_{z_c, i_x, i_y}^{MLMOr}   = \sum_{i = 1}^{N_{l}^{*}}  V_{z_c + (i-1) \, D_S, i_x, i_y}^{SLMOr} &&
\end{eqnarray}
where distance of $i$th layer to the coil is given by $z_c + (i-1) \, D_S$ and $N_{l}^{*}$ is found by  either maximizing the volumetric energy efficiency of EH device or maximizing the received power $P_h$ for single device unit. Number of layers is limited by $N_l \,   =  \, \mbox{min} \lbrace D_T^{max} \, / \, D_S, D_{max} \, / \,  D_S \rbrace$  where $D_T^{max} = \lambda_e \, / \, k_{\Theta}$ and $D_{max}$ is the application specific EH device height, e.g., bounded by the biological unit that the coil to be placed.   The number of layers  is optimized as follows:
\begin{equation}
\mbox{\textbf{max     }} P_h \,/ \,(z_c^{ a} \, + \,  w \, + \, N_{l}^{*}\,D_S) \mbox{\textbf{  s.t.  } } N_l^{*} \leq  N_l
\end{equation} 
where resulting device has the highest volumetric  EH efficiency computed with $Eff_h = P_h \, / \, V_r$ in ($W/m^3$) in a single coil and  device volume $V_r$ is approximated by $A_r \times h_r \approx A_r \times ( w + N_{l} \, D_S)$ for  $z_c^a \ll 1$. Device area values  ($A_r$) for the two schemes  are $A_C$  and $\mbox{max} \lbrace  A_C, (2\, N_g+1)^2 \, \Delta_g^2 \rbrace $ while $A_C$  equals to $\pi \, (N_s  \, + \,1)^2 \, \Delta_c^2$  and $(2 \, N_s  \, + \,1)^2 \, \Delta_c^2$, for Sch-1 and Sch-2, respectively.
 
\section{Experimental Challenges and Analysis of Modeling Accuracy}
\label{s5} 

  There is a set of challenges to be solved for realizing prototype. Fundamental idea, i.e., EH from SMMs on vibrating 2D planes (such as black phosphorus and transition metal dichalcogenides (TMDCs) including  molybdenum disulfide ($\mbox{MoS}_2$) in addition to graphene) and utilizing for MIC and WPT allow different architectures to be designed specific to the application. Therefore, specific analysis and experimental verification methods are necessary.  Next,  potential sources of error in modeling and realization are discussed. Then, finite element method (FEM) based simulation to improve accuracy for complicated  geometries and  importance of experimental measurements to improve analytical models are discussed. Issues regarding manufacturing and geometrical design, toxicology and operation in biological environments are discussed.
 
\subsection{ Modeling Accuracy Analysis}
  There are approximations in analytical models  requiring experimental feedback. Different types of resonators in literature have varying performances specific to each design as listed in \cite{mgr1}. Potential sources of error requiring experimental feedback are summarized as follows:
\begin{itemize}
\item Effect of SMMs on vibration should be experimentally analyzed in terms of  frequency shift where the model in \cite{mgr3} is utilized, i.e., $\vert \partial f \vert= \,f_0 \,\vert \partial m \vert/ \, (2 \,m )$.
\item Vibration properties in THz regime in terms of hybrid design and hole geometry  should be experimentally determined for an improved model for high frequencies. 
\item Placement of multiple SMMs in a small area on resonator  and the distribution of their easy axes should be experimentally modeled while perpendicular and homogeneous distribution is assumed based on advanced grafting methods in \cite{cr1, e4}. Effect of vibration on the distribution and total magnetic moment should be experimented. 
\item Effects of external magnetic field in arbitrary angles should be modeled in terms of the generated torque on SMMs and their effect on vibration. 
\item Receiver coil design should be optimized based on the detailed model in \cite{bg1} and modeling accuracy should be improved by simulation and experimental studies.
\item Steady state and transient response characteristics  should be experimentally measured to model non-linear and unplanned effects of the environment and hybrid design.
\item It is assumed that multiple layers of resonators are placed in an ordered manner by using advanced manufacturing methods such as carbon nanomembranes (CNMs) as discussed in \cite{bg3} for a similar multi-layer architecture with vibrating quantum dots (QDs). Modeling errors in terms of vibrational properties in such a design should be experimentally determined.
\end{itemize}

\subsection{ FEM Simulations and Experimental Verification}

 Theoretical analysis is based on experimentally verified models of resonators in \cite{mgr1, mgr3}, magnetic characteristics of graphene-SMM hybrid in \cite{e6, e3,  cr1} and inductive coupling of a magnetic particle with a circular coil in \cite{noncoupling}.   Furthermore, proposed models form the first step of any FEM simulation, i.e., mathematical modeling. In addition, we simplify device geometry with the architecture in Fig. \ref{Figure_4} and provide discretization methods of calculation in Fig. \ref{Figure_5}. Since the set-up is simple as composed of resonators with magnetic particles and circular coils,  numerical analysis without FEM based discretization is a good approximation compared with FEM simulations used for complicated systems. However, proposed designs should be simulated with FEM tools and experimentally measured to improve the analytical model and to accurately model nonlinear effects in terms of deflection, multi-layer design and interactions with SMMs.

Resonator is assumed to be  oscillating periodically without nonlinear terms at higher bands. Performance degradation with nonlinear carrier signals due to  mechanical effects, SMM mass, rotational or similar complicated mechanisms should be theoretically modeled, numerically simulated and experimentally  determined. Synchronization between  layers and performance of varying sets of excitation sources should be experimented. Storage of  harvested energy for  Mod-1  should be designed with high efficiency by analyzing  state of the art supercapacitive devices. Furthermore, methods should be developed to stack multiple layers with nanoscale precision. 
  
\subsection{Toxicology and Effects of Biological Environments}

Toxicology of graphene is discussed in detail in \cite{bg1} where various encapsulation methods and  studies on the effects of graphene on organs are discussed. Graphene is promising  as a bio-compatible device. Terbium compounds  have low to moderate toxicity, however effect of Tb$\mbox{Pc}_2$ SMM on cells should be analyzed experimentally in detail \cite{toxictb}. If the proposed device results in toxicity for \textit{in vivo} environments, then a suitable encapsulation to reduce the effects of toxicity without affecting the resonance and performance should be designed. However,  proposed resonators can be utilized for \textit{in vitro} studies,  e.g., microfluidic applications, with much more reliability. Effects of biological environment and  performance degradation due to eddy currents and coupling with the environment are discussed in detail in \cite{bg1} where isolating solutions and negligible effects of the environment are assumed.
 
\section{ Nanoscale and Biomedical Applications}
\label{s5b}

\begin{table}[t]
\centering
\caption{Numerical  Analysis Parameters}
\label{Table_3}
\footnotesize
\includegraphics[width=3.5in]{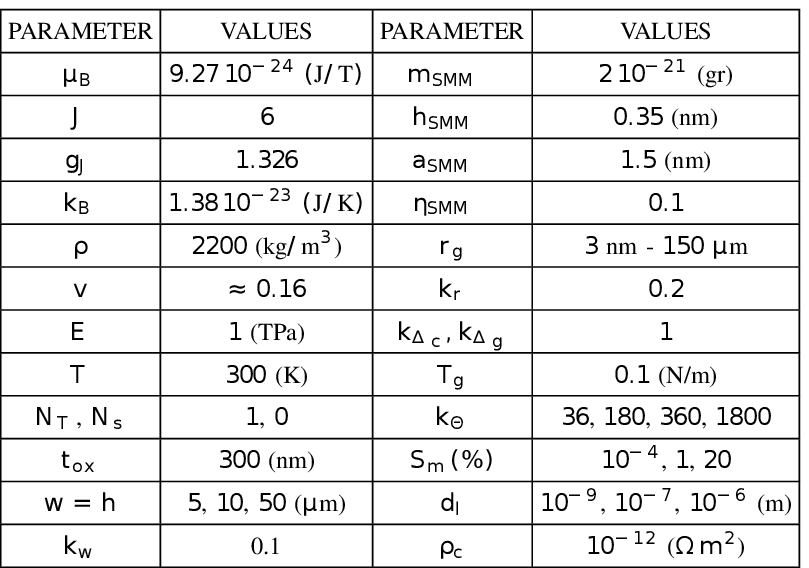}
\end{table}

Exploiting EH, signal carrier generation in  the KHz to THz band and modulation capabilities combined with  unique properties of graphene and SMMs  promise novel applications  for  in vivo  biomedical monitoring and sensing in diagnosis and treatment processes.  Graphene layers are  either resonated by external forces by harvesting vibration energy from biological forces or  by various internal mechanisms described in Table \ref{figtabmod}. Non-invasive and continuous EH capabilities combined with ultra-low weight, strong and nanoscale devices promise in-body and wearable applications.   Future nano-robotic transceivers can utilize EH  for autonomous operation while providing a simple method to modulate data with SWIPT shown in Fig. \ref{figtb1}(b) and Table \ref{Table_2}.

Cellular level tracking applications  such as cancer detection  for in vitro microfluidic systems can be designed with simple resonators improving  current optical methods based on fluorescence,   e.g., \cite{bg3}, or magnetic particle based tagging systems. MPI studies are improved by injecting a massive amount of bio-compatible graphene coils, creating waveguides and analyzing  induced currents with different frequencies.   Mechanisms to exploit proposed system architecture for MPI purposes are left as future works. Capability to generate oscillating magnetic fields, wideband frequency tuning, high signal amplitudes and important mechanical features such as lightweight design, durability and strength promise the utilization of SMMs for such purposes with the resolution determined by the resonator sizes. Graphene is lipophilic making it possible to be used to monitor cellular processes \cite{gref1}.  In addition, THz signal carrier generation can be utilized for non-contact feeding of nanoscale coils by reducing the effects of contact resistances and simplifying the  design  in \cite{bg1}. Proposed device  promises on-chip THz applications  \cite{bg1}. 

\section{Numerical   Analyses }
\label{s6}

\begin{figure}[!t]
\begin{center}
\includegraphics[width=3.5in]{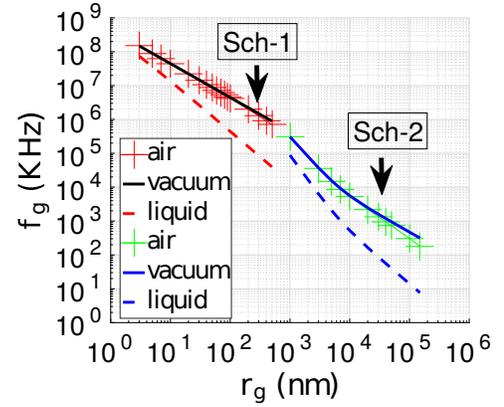} 
\caption{ Resonance frequencies for varying $r_g$  of  Sch-1 and Sch-2 type devices.}
\label{Figure_6}
\end{center}
\end{figure}

\begin{figure*}[!t]
\begin{center}
\includegraphics[width=7in]{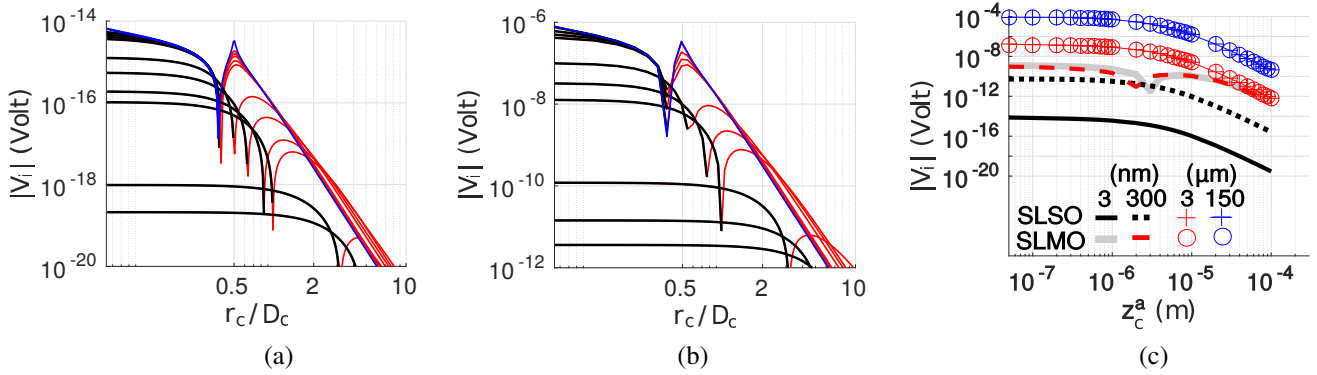}   
\caption{ SLSO performance  in air medium for  varying $r_c$   with   $S_m = 20\%$ and $w = 5 \, \mu$m coil size due to (a) varying $z_c^a$ between 50 nm and 60 $\mu$m   in Sch-1  with  $r_g = 3$ nm and $h_g = 0.335$ nm (SLG), and (b) varying $z_c^a$ between 50 nm and 100 $\mu$m  in Sch-2 with $r_g = 10 \, \mu$m and $h_g = 30$ nm (MLG). (c) The performance comparison of SLSO and SLMO devices in air  where  $w = 5 \, \mu$m, optimized $r_c$ and  $S_m = 20\%$ for  varying $z_c^a$  and  $r_g$ due to  Sch-1 device of $h_g = 0.335$ nm (SLG) with $r_g = 3$ nm and $300$  nm  and  Sch-2 device  of $h_g = 30$ nm (MLG) resonators with $r_g = 3 \, \mu$m and $150 \, \mu$m.}
\label{Figure_7}
\end{center}
\end{figure*}

Sch-1 and Sch-2 devices are simulated for varying dimensions of coils and resonator radius.   Analysis parameters are summarized in Table \ref{Table_3}.  Resonator radius $r_g$ is chosen  between $3$ nm  and $150$ $\mu$m with $f_0$   between  several KHz  and hundreds of GHz. Resonator frequencies are shown in Fig. \ref{Figure_6}  based on detailed model in  Section \ref{s2} reaching mmWave frequencies of $\approx 150$ GHz for SLG device of the type Sch-1 while covering ultrasound and acoustic frequencies for  MLG device of the type Sch-2.  Frequency spectrum can be extended with  experiments promising THz frequencies theoretically in \cite{mn1} for high performance  MIC and WPT as discussed in \cite{bg1}.   Analyzed acoustic frequencies are validated by theoretical and experimental studies  \cite{mgr1}. 
  
Resonator thickness  $h_g$ is assumed to be either $0.335$ nm (SLG) or $30$ nm (MLG)  in compatible with experimental and theoretical studies with SLG and MLG  \cite{mgr1}. Properties of resonators are calculated in  Section \ref{s2} while utilizing the  constants defined in Table \ref{Table_3}  for $\mbox{TbPc}_2$ SMM and graphene, i.e., $\mu_B$, $g_J$, $J$, $\rho$, $v$, $E$, $T_g$, $t_{ox}$ and room temperature $T = 300$ K.   $\eta_{SMM} = 0.1$ results in a frequency shift of $5\%$ at most where $m_{SMM}$, $h_{SMM}$ and  $a_{SMM}$ are the weight,  height and   side length of single $\mbox{TbPc}_2$ molecule, respectively \cite{tbsize}. They are placed on resonator center with small $k_r = 0.2$.   Independent and separated resonators  are realized with $k_{\Delta_g} = 1$.

\begin{figure*}[!t]
\begin{center} 
\includegraphics[width=7in]{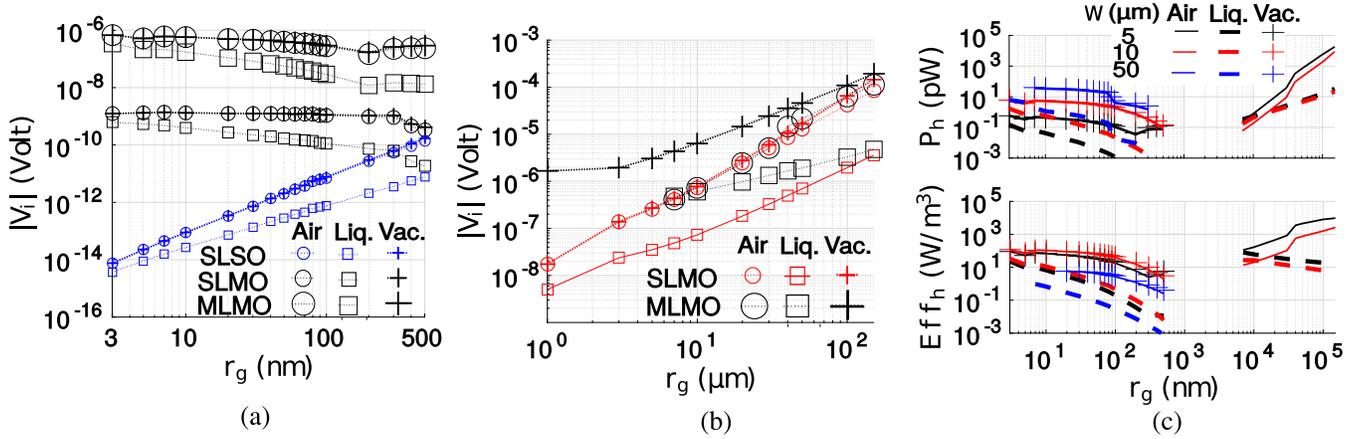}
\caption{ SLSO, SLMO and MLMO performances with $w = 5 \, \mu$m size single coil for varying $r_g$ and medium  where $z_c^a = z_{min}$, $d_l = 1$ nm, $k_{\theta} = 360$, $S_m = 20\%$ and  $D_{max} = 50 \, \mu$m showing (a) Sch-1 and (b) Sch-2 type device performances, and  (c) their $P_h$ and $Eff_h$ MLMO  performances for varying $w$ and the medium.}
\label{Figure_8}
\end{center}
\end{figure*}

MLMO performance is simulated for varying device parameters and strain levels. $d_l$ is chosen between $1$ nm and $1 \, \mu$m while both the compact layering and more relaxed layering mechanisms are considered, respectively. $k_{\Theta}$ is chosen between $36$ and $1800$  while analyzing lower thickness and more synchronized layers as $k_{\Theta}$ increases. $S_m$ is  analyzed  for high level of strain of $20\%$ and lower levels of $1\%$ and $10^{-4}\%$ to  analyze  harvesting from low power excitations. The noise in the receiver is assumed to be of thermal type with noise spectral density given by $N_{Th} = \, k_B \, T$ \cite{bg1}.

Coils are assumed to be made of suspended MLG with intercalation as discussed in detail in \cite{bg1}. Detailed theoretical modeling of the resistive and capacitive effects of the coil are summarized in the Appendix and in \cite{bg1} while more accurate modeling with experimental verification is an open issue. Three different coil dimensions are designed for   microscale monitoring,  MIC and  EH purposes, i.e., $w = h = 5, \, 10$ and $50$ $\mu$m, where $a_i = 2 \, \times \, k_w \, \times \, w = w \, / \, 5$ and $a_e =a_i \, + \, 2 \, \times \, w$ with    $k_w = 0.1$ for compact dimensions.  Single-turn coil ($N_T = 1$)  is chosen to   have  simplification and manufacturing for massive production and placement into biological environments. The cases with $w \geq 10 \,\times \, r_g$ is realized with Sch-1 device type to place resonators  more compactly   with less computational complexity to calculate harvested energy. Otherwise,  Sch-2 is utilized with  SO  centers on a grid. The number of coils is chosen as   one with $N_s = 0$ and    $k_{\Delta_c} = 1$  to simplify the analysis and to  have compact and simple devices. Coils  have $\vect{n}_c = [0 \, \, 0 \, \, 1]$ for EH purposes at  $[x_c\, \, 0 \,\, z_c]$ resulting in both vertical displacement $z_c^a \equiv z_c - \, h \,/ \, 2$ (neglecting oxide thickness $t_{ox}$)  and radial displacement $r_c = \sqrt{x_c^2 + y_c^2} = x_c$.  

Contact resistivity of $\rho_c = 10^{-12} $  ($\Omega \, m^2$) is assumed based on the promising developments in nanotechnological architectures \cite{bg1}. The resulting resistance value $R_{eff}$ for the specific example in Section \ref{s6s2} is $\approx 22.3 \, \mbox{m}\Omega$ with contact resistance limited regime of $R_c \approx 20 \, \mbox{m}\Omega$ for the compact coil with $w = h = 10 \, \mu$m at $f_0 = 87$ KHz.  It requires experiments for improving accuracy and verification of the theoretical model  \cite{bg1}. All-graphene architectures with smaller contact resistances promise higher performances.

Three different mediums are considered, i.e., air, liquid or vacuum, with different damping effects to the frequency and different wavelengths of the excitation signal.  Two different excitation signals, i.e., acoustic and the one with the speed of light such as optical, are assumed in addition to the electrostatic excitation as shown in Table \ref{figtabmod}. Velocity of the excitation signal is assumed to be $c = 3 \times 10^8$ m/s for Sch-1 devices having the frequencies larger than $10$ MHz while the acoustic velocity is assumed to be $343$ m/s and $1540$ m/s in air and liquid medium, respectively.   In the next sections,   EH and MIC performances are  analyzed.  

  \subsection{Wireless Energy Harvesting Performances}
\label{s6s1}

In the following, SLSO, SLMO and MLMO  EH performances are simulated for varying $r_g$, $r_c$, $z_c^a \,  > z_{min} = 50$ nm, strain level $S_m$, $d_l$ and $k_{\Theta}$, and medium.

\subsubsection{Single Layer Resonator Performances}
\label{s6s1s1}

Device performances of SLSO Sch-1 and Sch-2 devices are shown for varying radial and vertical distances with $w = 5 \, \mu$m   size single coil  in  Figs. \ref{Figure_7}(a) and (b), respectively. It is observed that  radial performance drops significantly as   $r_c$ is larger than several times  $D_c$  ($D_c$ defined as $ a_e$ for side length of the device with $N_s = 0$ instead of $\Delta_c$) for both Sch-1 and Sch-2. Furthermore, as  $r_c$ is changed, induced voltage changes sign as shown with different colors. Since  SMMs are assumed to reside at the center of the resonator for Sch-2, i.e., for $r_g > r_c$,  receiver diversity cannot be utilized efficiently to harvest more energy by placing more coils inside  single resonator area. 

 SLMO performance for $z_c^a$ at an optimized $r_c$ for the maximum induced voltage is shown in Fig. \ref{Figure_7}(c). As the number of resonators  ($N_R$) increases for Sch-1,  performance increases significantly reaching even $\approx 10^5$ improvement for $r_g = 3$ nm resonator and $w =  h = 5 \, \mu $m coil dimensions.  As the distance is comparable to the several times the coil diameter,  the voltage drops with an attenuation constant of almost $\alpha = 4$  in the medium range  with    $\alpha = 8$ for power, i.e., $P_h  \propto \vert V_i\vert^2$ as discussed  in the Appendix.  Sch-1  achieves similar levels of the performance for the same total area as shown in   Figs.  \ref{Figure_7}(c)  and Fig. \ref{Figure_8}(a).  If it is assumed that the distribution  of the radial distance between the coil and resonators are similar due to the significant  $N_R$ for large coil area  and $\Phi_{\widetilde{z}, u}^n$ shows similar values with normalized magnetic moment in (\ref{eq2})  then,  $V_i$ with SLMO devices for Sch-1  is simplified as approximately proportional to the following: 
\begin{eqnarray}
\frac{\vert V_i  \vert }{ \vert \vect{m} \vert} \propto   N_{R} \, N_{SMM}  \, A \, \omega_0 \, \hspace{0.9in}& \\
\label{slrp_eq1}
 \propto   \frac{ \Delta_c^2}{r_g^2} \, \frac{ \eta_{SMM} \,   r_g^2 \, h_g \, \rho}{m_{SMM}} \,   (\sqrt{S_m}\, r_g )  \, \frac{1}{r_g} \, &\\
\label{slrp_eq2}
 =   \frac{\Delta_c^2 \,  \eta_{SMM} \,h_g \, \rho \,    \sqrt{S_m}}{m_{SMM} } \hspace{0.65in} &
\label{slrp_eq3}
\end{eqnarray}
where  the radius of total area is proportional to $\Delta_c$, $A \propto \sqrt{S_m} r_g$  and $\omega_0 \propto 1 \, / \, r_g$ for SLG.  

\subsubsection{Multi-layer Resonator Performances}
\label{s6s1s2}

In Figs. \ref{Figure_8}(a) and (b),  comparison  among SLSO, SLMO and MLMO devices are shown in various medium of air, liquid and vacuum. The other  parameters are  $d_l = 1$ nm, $z_c^a = 50$ nm, $k_{\theta} = 360$ for highly synchronized resonating layers, $S_m = 20\%$ and  $D_{max} = 50 \, \mu$m  suitable for  microscale applications.   As shown in Fig. \ref{Figure_8}(a), MLMO significantly improves SLMO  performance for all  medium types reaching $\approx 10^3$ times improvement while $\approx 10^8$ for SLSO.  A similar improvement is observed for large resonator sizes with the improvement vanishing as the number of layers decreases  (increasing $r_g$) as shown in Fig. \ref{Figure_8}(b). Sch-2 performances are higher than Sch-1 replacing  $\Delta_c^2$ in (\ref{slrp_eq3}) with increasing $r_g^2$. Furthermore, liquid  leads to lower operating frequency and performance.

In Fig. \ref{Figure_8}(c), it is observed that tens of pW level powers are harvested including mmWave frequencies around $150$ GHz while tens of nWs power can be harvested  at acoustic frequencies with larger resonators of tens of micrometer radius  (with the corresponding frequencies in Fig. \ref{Figure_6}).   $Eff_h$ ($W/m^3$) reaches several hundreds at mmWave frequencies to orders of $10^4$  at acoustic and ultrasound frequencies with a significant  nanoscale EH  performance. These results are comparable with state of the art triboelectric harvesters with efficiencies on the order of $10^4$ ($W/m^3$) in \cite{tribo} but with much smaller sizes to be embedded in nano-biological units. Simplicity of manufacturing, planar structure and MI based applications are other advantages.  An upper bound of  EH is provided for electromagnetic harvesters with $1.9 \, 10^{-6}\, V_r^2  f_0^2$ in \cite{e13} where $V_r$ is in $\mbox{cm}^3$ and $f_0$ is in Hz. This results in $17.5$ nW for  $r_g = 100 \, \mu$m, $w = h = 5 \, \mu$m,  $z_c^a = 50$ nm, $f_0 \approx 300$ KHz and in air medium where the dimensions of the optimum efficiency device is $\approx 200 \, \mu m \times 200 \, \mu m \times  8 \, \mu m$. The proposed device  has the maximum peak power of $\approx 5.7$ nW as shown in Fig. \ref{Figure_8}(c)  comparable with the maximum state of the art generators although  exact  dimensions should  include storage units such as   supercapacitors to make a more fair comparison.

\begin{figure*}[!t]
\begin{center}
\includegraphics[width=7in]{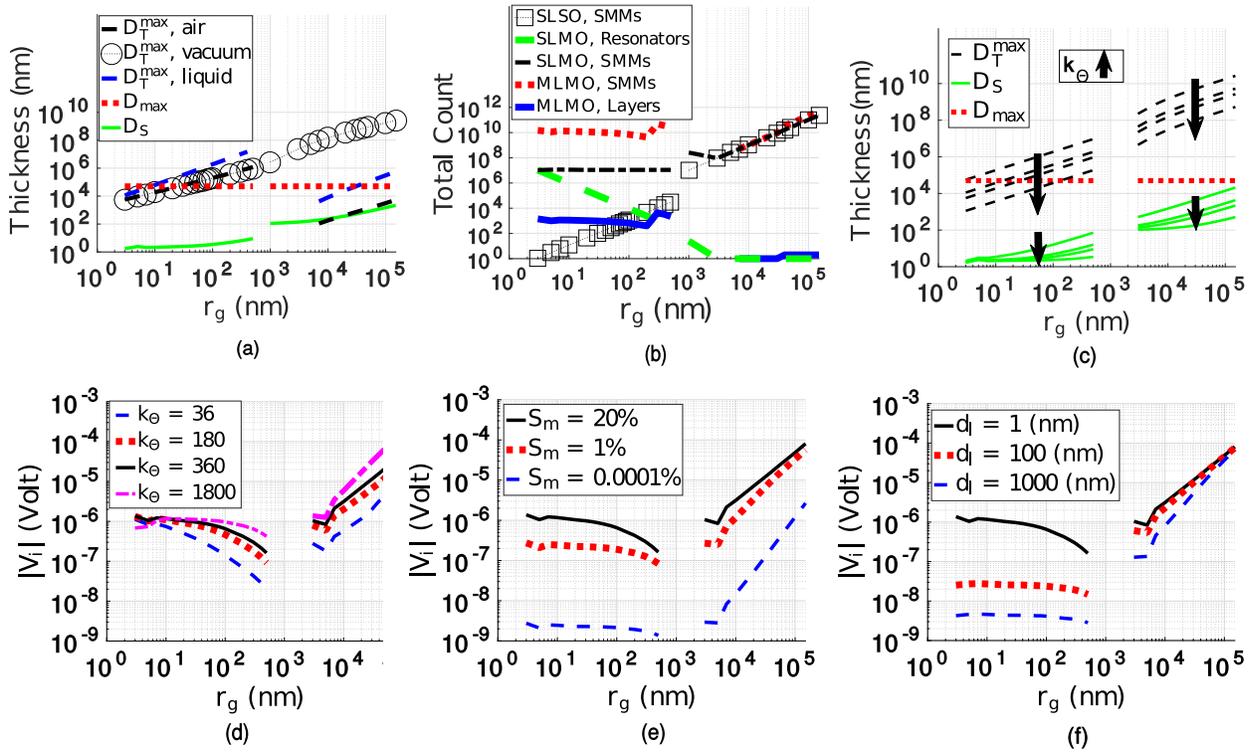}
\caption{(a)  $D_{T}^{max}$, $D_{max}$ and single layer thickness $D_S$ for  varying medium and $r_g$, and (b) total counts of SMMs, layers and resonators in air for varying $r_g$ where SLSO, SLMO and MLMO devices have $w = 5 \, \mu$m, $z_c^a = z_{min}$, $d_l = 1$ nm, $k_{\theta} = 360$, $S_m = 20\%$ and  $D_{max} = 50 \, \mu$m.  (c) Bounding and single layer thicknesses, and (d) $\vert V_i \vert$  in vacuum with $w = 10 \, \mu$m,  $d_{l} = 1$ nm, $S_{m} = 20\%$ and $z_c^a = z_{min}$ for varying $k_{\Theta}$ and $r_g$.  $\vert V_i \vert$   in vacuum with $w = 10 \, \mu$m,  $k_{\Theta} = 360$ and $z_c^a = z_{min}$ (e) for varying $S_{m}$ and $r_g$ with $d_{l} = 1$ nm and, (f) for varying  $d_{l}$  and $r_g$ with $S_{m} = 20 \%$.}
\label{Figure_9}
\end{center}
\end{figure*}
  
$D_T^{max}$, $D_{max}$ and $D_S$, and the numbers of SMMs, resonators and  layers in SLSO, SLMO and MLMO devices are shown in Figs. \ref{Figure_9}(a) and (b), respectively.   $D_T^{max}$ of Sch-1   with small $r_g$ is similar for  varying medium as shown in Fig. \ref{Figure_9}(a). The number of layers is increased for Sch-2  with respect to $\mbox{min} \lbrace D_T^{max} \, / \, D_S, D_{max} \, / \,  D_S \rbrace$ in water and in air with increasing $D_T^{max}$.   $D_S$ values are smaller than $\approx 20 \, \mu$m  suitable to be placed into  nanoscale units.   In addition, number of resonators with the smallest $r_g$  reaches $\approx 10^7$ for Sch-1 while it succeeds to have an approximately constant number of SMMs and layers for small $r_g$ due to dependency in (\ref{slrp_eq3}) and minor increase in $D_S$ in Fig. \ref{Figure_9}(b). Number of SMMs reaches to the orders of $10^{10}$ and $10^{11}$ for Sch-1 and Sch-2, respectively, while the number of layers is smaller than $\approx 2000$ and $2$ for Sch-1 and Sch-2 devices, respectively, in air. 
 
\subsubsection{Effects of Phase Synchronization, Interlayer Distance and Strain}
\label{s6s1s3}
  
The relation between  $D_T^{max}$ and  $D_S$, and $V_i$  performance for varying  $k_{\Theta}$ are shown in Figs. \ref{Figure_9}(c) and (d), respectively, for $w = 10 \, \mu$m coil, $z_c^a = z_{min}$, $S_{m} = 20\%$ and $d_{l} = 1$ nm in vacuum. It is observed that as $k_{\Theta}$ increases $D_T^{max}$ decreases while the decrease changes the total number of layers more for longer resonator radius. However,   total thickness is     becoming lower than $D_{max}$ limiting the number of layers   for small $r_g$ as shown in Fig. \ref{Figure_9}(c). Performance improvement is observed for both Sch-1 and Sch-2 bounded by  $\mbox{min} \lbrace D_T^{max} \, / \, D_S, D_{max} \, / \,  D_S \rbrace$    as shown in Fig. \ref{Figure_9}(d) while  improvement is more clearly observed for larger radius Sch-1 resonators.  Effect of $d_l$ is similar to $k_{\Theta}$ changing the number of layers and  performance especially on Sch-1 with SLG layers  as shown in Fig. \ref{Figure_9}(f). However,  effect of strain is different such that  increasing number of layers due to low strain cannot overwhelm the reduction due to lower vibration amplitude  as shown in Fig. \ref{Figure_9}(e).

\subsection{Wireless Communications Channel Performance}
\label{s6s2}

\begin{figure}[!t]
\begin{center}
\includegraphics[width=3.5in]{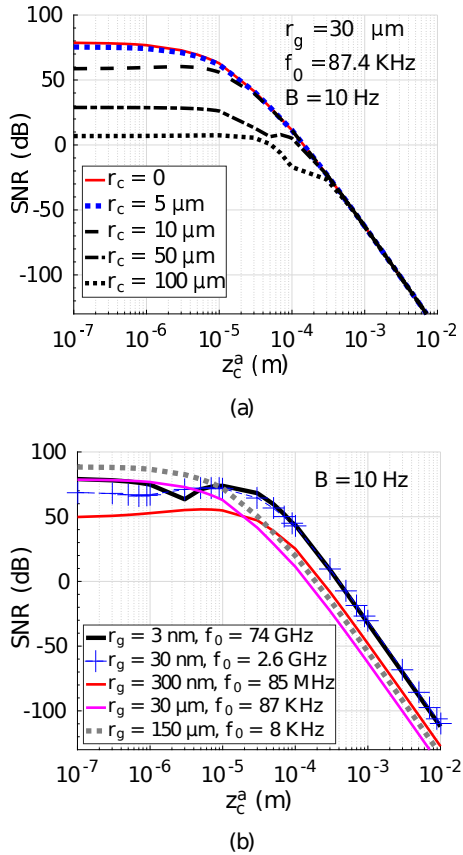} 
\caption{ SNR   performances of MLMO devices in liquid (a) for varying $r_c$  and $z_c^a$ with  $r_g = \, 30 \, \mu$m, and (b) for varying $r_g$ and $z_c^a$ at $r_c = 0$, where  $w= 10 \, \mu$m, $S_{m} = 20\%$, $d_{l} = 1$ nm and $k_{\Theta} = 360$.}
\label{Figure_10}
\end{center}
\end{figure}

Received power in a single coil due to MLMO resonators is expressed as $ P_h = \vert V_{i} \vert^2 \, / \, (8 \, R_{eff})$ as discussed in the Appendix. If we assume the receiver operates in the thermal noise limited regime and a low data rate monitoring is realized by detecting the oscillation frequency of SMMs with tuned coil receivers, then SNR  at the frequency $\omega$ is expressed as $ SNR =  P_h \, / \, N_{Th}  B$ where  $B$  denotes ultra-narrow bandwidth of the resonator. The channel geometry  provided in Fig. \ref{Figure_2}(b) is  analyzed  for varying $r_c$, $z_c^a$ and $r_g$ where $ B  = 10 $ Hz  and $w = 10 \, \mu$m.   The ultra-narrow bandwidth allows low data rate communications while the limit to lower the bandwidth depends on the excitation source or the mechanical effects. If  frequency tagging is utilized, then different resonators are placed  or attached with target units, e.g., cells or nano-robots,  and they are tracked based on the detection of the frequency at distances of several millimeters providing a novel framework for future nanoscale  MPI  or tracking applications.  

 Performances without angular misalignments of  MLMO devices for  varying vertical and radial distances are shown in Fig. \ref{Figure_10}(a) for liquid medium with $w = 10 \,\mu$m, $r_g = 30 \, \mu$m, $f_0  \approx 87.4 $ KHz and $B = 10$ Hz. Performance decrease for  varying $z_c^a$ with small $r_c$  is limited to $3$ dB for  $r_c \leq 5 \, \mu$m while the drop at small  $z_c^a$   is significantly high for larger  $r_c$. It is possible to have zero SNR level for hundreds of micrometer vertical distances with smaller effect of varying $r_c$. In \cite{bg1}, the effects of angular misalignments on MI waveguide channels   extending ranges significantly  are discussed emphasizing the robustness to small angular disorientations less than $\pi/3$ for radial displacements less than  $2.5$ times the coil radius.  Numerical analysis for angular misalignments is left as a future work in a more complicated relaying waveguide scheme.

Performance for varying $r_g$ is shown in Fig. \ref{Figure_10}(b) with the same medium, coil size and bandwidth parameters. SNR drop is $\approx 80$ dB per decade while it is possible to receive larger than zero SNR at distances reaching $400 \, \mu$m for mmWave frequencies and hundreds of micrometer distances for acoustic and ultrasound frequencies. 

As a future work, the effect of passive coils within a waveguide structure will be analyzed to improve the range to much longer distances including macroscale ranges as discussed in \cite{bg1}. As the number of  units attached with coils increases, they perform as waveguides for the resonating SMMs potentially allowing a \textit{waveguide relaying network} to carry the signals to the external receivers. Resonators can be embedded into nanoscale communication networks and biological units, e.g., cells or similar microscale volumes  to track or identify them at millimeter ranges. Waveguide scheme  provides a novel form of MPI application with cellular resolution. A low cost  architecture can be achieved for simultaneous multi-cell tracking  applications. Performance analyses of  waveguides, angular orientations and densities of resonators are open issues.

\section{Conclusion}
\label{conclusion}

In this article, graphene nanoscale resonators are combined with SMMs to realize  nanoscale EH, MIC and  WPT simultaneously by exploiting unique   mechanical and geometrical advantages of graphene with  high and stable orbital magnetic moment of $\mbox{TbPc}_2$ SMM grafted on graphene.  Low complexity mechanical design supports novel modulation methods achieving SWIPT with real-time carrier signal generation  in the KHz to THz band.  Numerical   analysis presents tens of nanowatts power EH and efficiencies of $10^4 \, W/m^3$ in acoustic and ultrasound frequencies as nanoscale and high frequency alternatives to state of the art electromagnetic vibrational EH devices.    Challenges to realize system prototype are discussed. Proposed system design promises applications in the near future such as nanoscale tagging of biological structures, biomedical sensing and communications,  EH and modulation for nano-robotic transceivers and  MPI systems.

\appendix

\section{Equivalent Circuit of the Graphene Coil}
\label{app3}

\begin{figure}[!t]
\begin{center}
\includegraphics[width=3.5in]{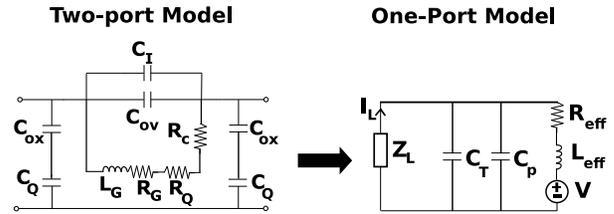} 
\caption{Two-port model of graphene OC inductor and circuit theoretical modeling of WG  with one-port coil models \cite{bg1}.}
\label{Figure_11}
\end{center}
\end{figure}
 
The equivalent circuit model is defined in detail in \cite{bg1}. $C_{ov} = \epsilon_{r, ox} \, \epsilon_0 \, A_{o} \, / \, h_{ov}$ where  $A_o = N_T \, w^2$, $N_T$ is the number of turns, $h_{ov} \equiv t_{ox}  \, / 2$, $\epsilon_{0} = 8.854 \times 10^{-12}$ (F/m) and $\epsilon_{r, ox} = 3.9$ are absolute vacuum and relative oxide permittivity values, respectively.  Furthermore,  $C_{ox}$ is approximated with distributed capacitance modeling for graphene while $C_I$ is neglected due to $N_T = 1$ \cite{bg1}.  $C_Q$ equals to $ N_{ch} \,  N_{L} \, 4 \, q^2 \, / \, (h_p\, v_F)$    where $h_p = 6.626 \times 10^{-34}$ (J $\times$  s) is Planck's constant, $q$ is electron charge,   $N_{L} = h \, / \, s_{L}$ is the number of  single layers,   $ s_{L} =  0.575$ nm, $N_{ch}$ is the number of conducting channels in one layer given by $\sum_{j=0}^{n_c}( 1 + e^{(E_j - E_F) \, / \, k_B T})^{-1} \, + \, \sum_{j=0}^{n_v}(1 + e^{(E_j + E_F) \, / \, k_B T})^{-1}$ where $E_j = j \, h_p \, v_F \, / \, 2 \, w$ assuming metallic graphene, $v_F \approx 8 \times 10^5$ m/s is the  Fermi velocity and  $E_F = 0.6 \, eV$.
Parameters shown in Fig. \ref{Figure_11} are obtained as $C_{p} \,  =  \, C_{I} \, + C_{ov} \, + (1/C_{ox} + 1/C_Q)^{-1}$ and $
R_{eff} \,  =   R(\omega) \, + \, R_Q \, + \, R_c$  where  $\omega \, = \, 2 \, \pi \, f$,  $L_{eff} =  L(\omega)$, $R(\omega)$ and $L(\omega)$ are  equivalent resistance and inductance, respectively,   $R_c= 2 \, \rho_c \,  / \, (w \, h)$, $\rho_c$ is the contact resistivity in $(\Omega \, m^2)$ and $R_{Q} =  h_{p} \, / \, (2 \, q^2 \, N_{ch} \,  N_{L})$  \cite{bg1}. Then, the tuning capacitor $C_T $ and the matched load  $Z_L = R_L \, + \, \jmath \, X_L$ result in  $ P_h = \vert V_{p} \vert^2 \, / \, (8 \, R_{eff})$ at  $\omega$ and $V_p$   equals to $V_{z_c, r_c}^{MLMOc}$ and $V_{ z_c, i_x, i_y}^{MLMOr}$ for Sch-1 and Sch-2, respectively.

\section*{Acknowledgement}

The authors thank to Dr. Gorkem Memisoglu  and the support of Vestel Electronics.

\bibliographystyle{IEEETrantr}

\end{document}